\documentclass[12pt]{article}

\usepackage{graphicx}
\usepackage{amssymb}
\usepackage{amsmath}
\usepackage{dsfont}
\usepackage{bm}
\usepackage{mathrsfs}

\begin{document}
\title{Quantum optics of spatial transformation media}
\author{Ulf Leonhardt and Thomas G. Philbin\\
School of Physics and Astronomy, University of St Andrews,\\
North Haugh, St Andrews KY16 9SS, Scotland
}
\date{\today}
\maketitle
\begin{abstract}
Transformation media 
are at the heart of
invisibility devices, perfect lenses and 
artificial black holes.
In this paper, 
we consider their quantum theory. 
We show how transformation media
map quantum electromagnetism in physical space
to QED in empty flat space.\\

\end{abstract}

\newpage

\section{Introduction}

Transformation media \cite{PSS,LeoPhil}
map electromagnetic fields in physical space
to the electromagnetism of empty flat space.
Such media are at the heart of macroscopic
invisibility devices 
\cite{PSS,LeoPhil,Schurig,LeoConform,Hendi,Cai};
but also perfect lenses \cite{Pendry,Veselago}
and electromagnetic analogs of the event horizon
\cite{LeoPhil,LeoReview,SchUh}  
are manifestations of transformation media \cite{LeoPhil}.
A transformation medium performs an active coordinate transformation: electromagnetism in physical space, 
including the effect of the medium, 
is equivalent to electromagnetism in transformed coordinates 
where space appears to be empty;
the sole function of the device is to facilitate a coordinate transformation.
Figures 1-4 show a gallery of spatial transformation media
and illustrate their effects on electromagnetic waves. 

\begin{figure}[h]
\begin{center}
\includegraphics[width=30.0pc]{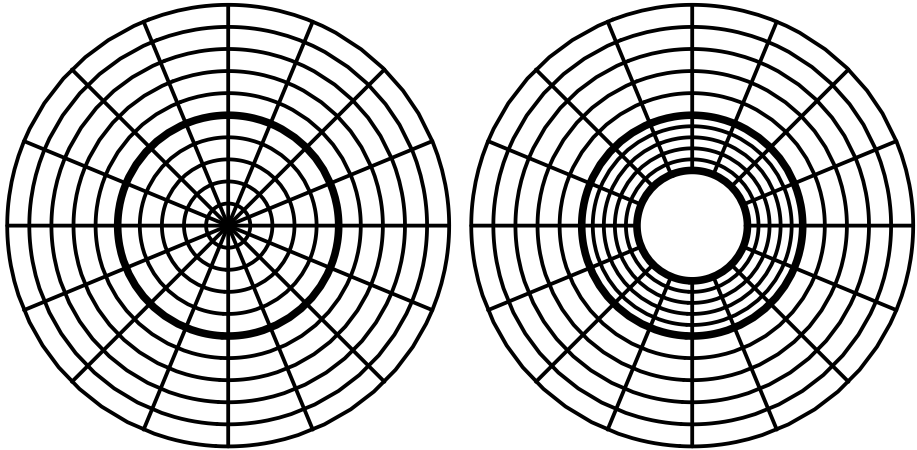}
\caption{
\small{
Transformation media implement coordinate transformations.
The right figure shows an orthogonal grid of coordinates
in physical space that is transformed to the grid on the left figure.
The physical coordinates enclose a hole that is made invisible
by the transformation.
Consequently, a medium that performs this transformation 
acts as an invisibility device.  
The case illustrated in the figure corresponds to 
the transformation  \cite{LeoPhil}
$r=R_1 + r'(R_2-R_1)/R_2$ in cylindrical coordinates
where the prime refers to the radius in transformed space.
The region with radius $R_1$ is invisible; $R_2-R_1$
describes the thickness of the cloaking layer.
}
\label{fig:spider}
}
\end{center}
\end{figure}

\begin{figure}[h]
\begin{center}
\includegraphics[width=16.0pc]{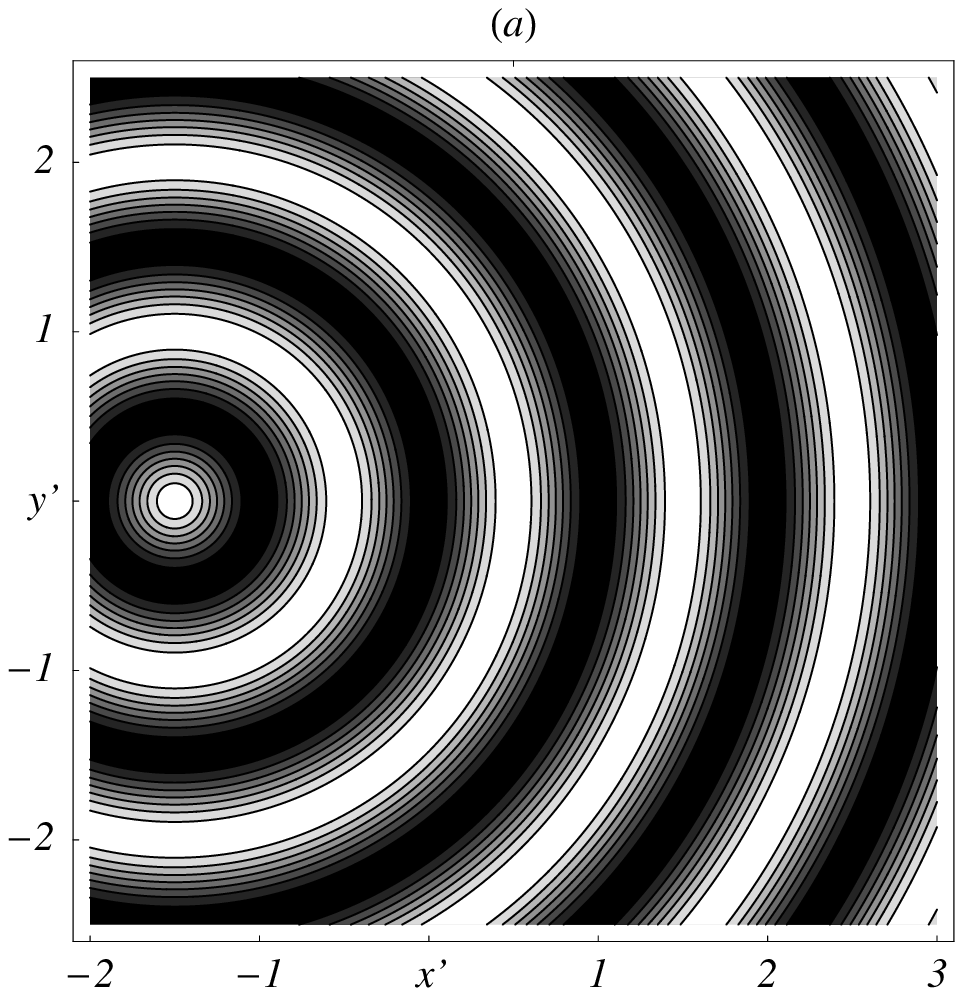}
\includegraphics[width=16.0pc]{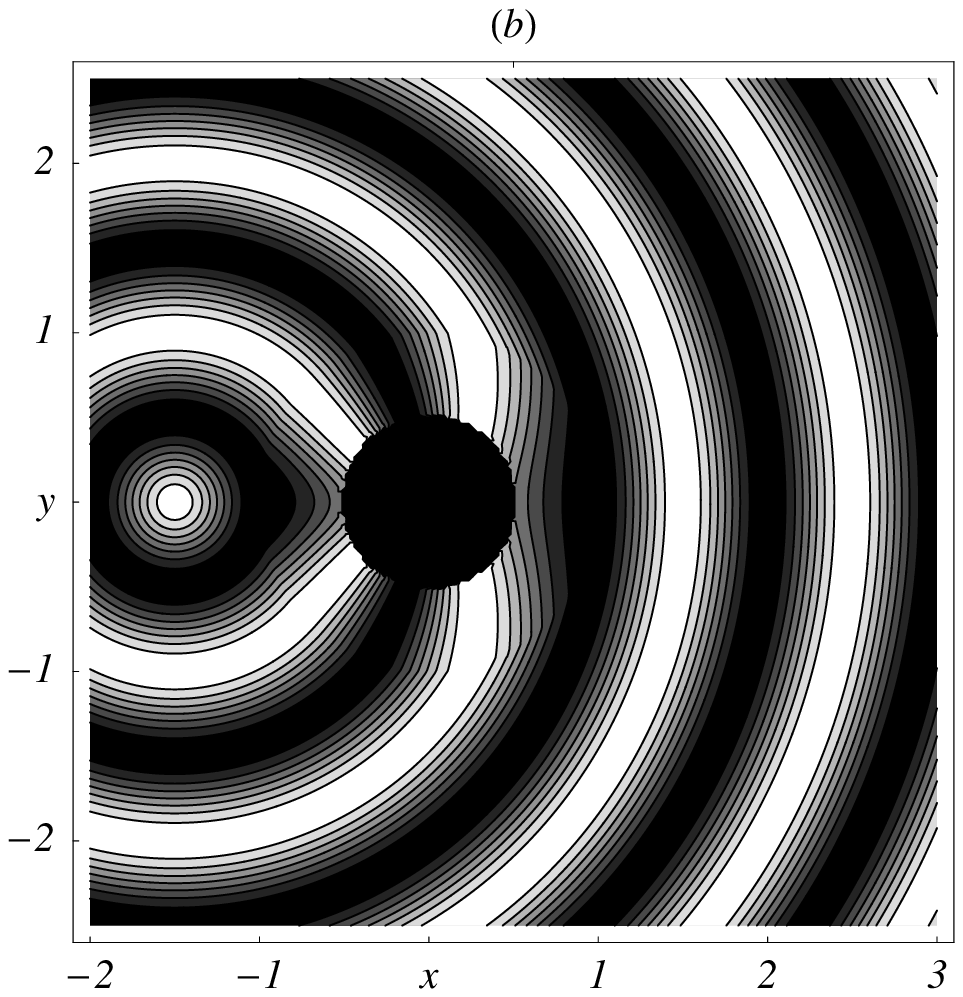}
\caption{
\small{
Invisibility device.
The device transforms 
waves emitted in physical space as if they propagate 
through empty space in transformed coordinates.
(a) Transformed space.
The figure shows the propagation of a
spherical wave emitted at the point $(-1.5,0,0)$.
(b) Physical space.
The invisibility device 
turns the wave of figure (a) into physical space
by the coordinate transformation illustrated in Fig.\ \ref{fig:spider}.
The cloaking layer with radii $R_1=0.5$ and $R_2=1.0$
deforms electromagnetic waves such that they
propagate around the invisible region
and leave without carrying any trace of the interior of the device.}
\label{fig:waves}
}
\end{center}
\end{figure}

\begin{figure}[h]
\begin{center}
\includegraphics[width=16.0pc]{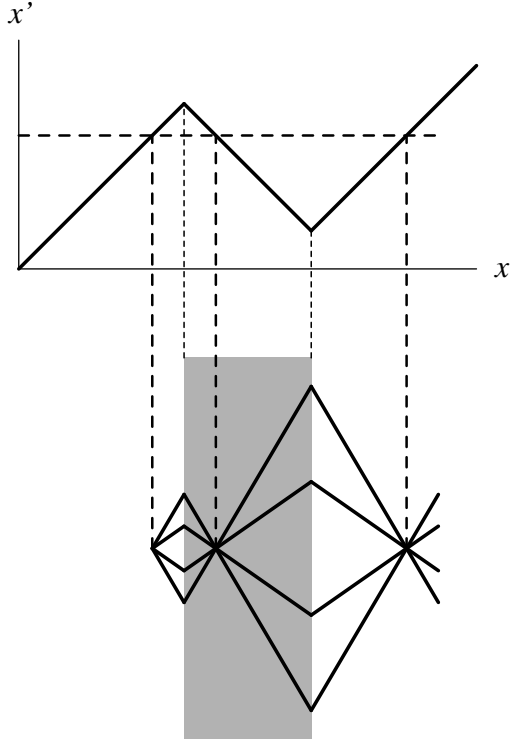}
\caption{
\small{
Perfect lens.
Negatively refracting perfect lenses employ transformation media.
The top figure illustrates the coordinate transformation (\ref{eq:left})
from the physical $x$ axis to $x'$,
the lower figure shows the corresponding device.
The inverse transformation from $x'$ to $x$ is either triple
or single-valued.
The triple-valued segment on the physical $x$ axis 
corresponds to the focal region of the lens: any source point
has two images, one inside the lens and one on the other side.
Since the device facilitates an exact coordinate transformation,
the images are perfect with a resolution below the
normal diffraction limit \cite{BornWolf}: 
the lens is perfect \cite{Pendry}.
In the device,
the transformation changes 
right-handed into left-handed coordinates.
Consequently, the medium employed here is left-handed, 
with negative refraction \cite{Veselago}.
}
\label{fig:lens}
}
\end{center}
\end{figure}

\begin{figure}[h]
\begin{center}
\includegraphics[width=16.0pc]{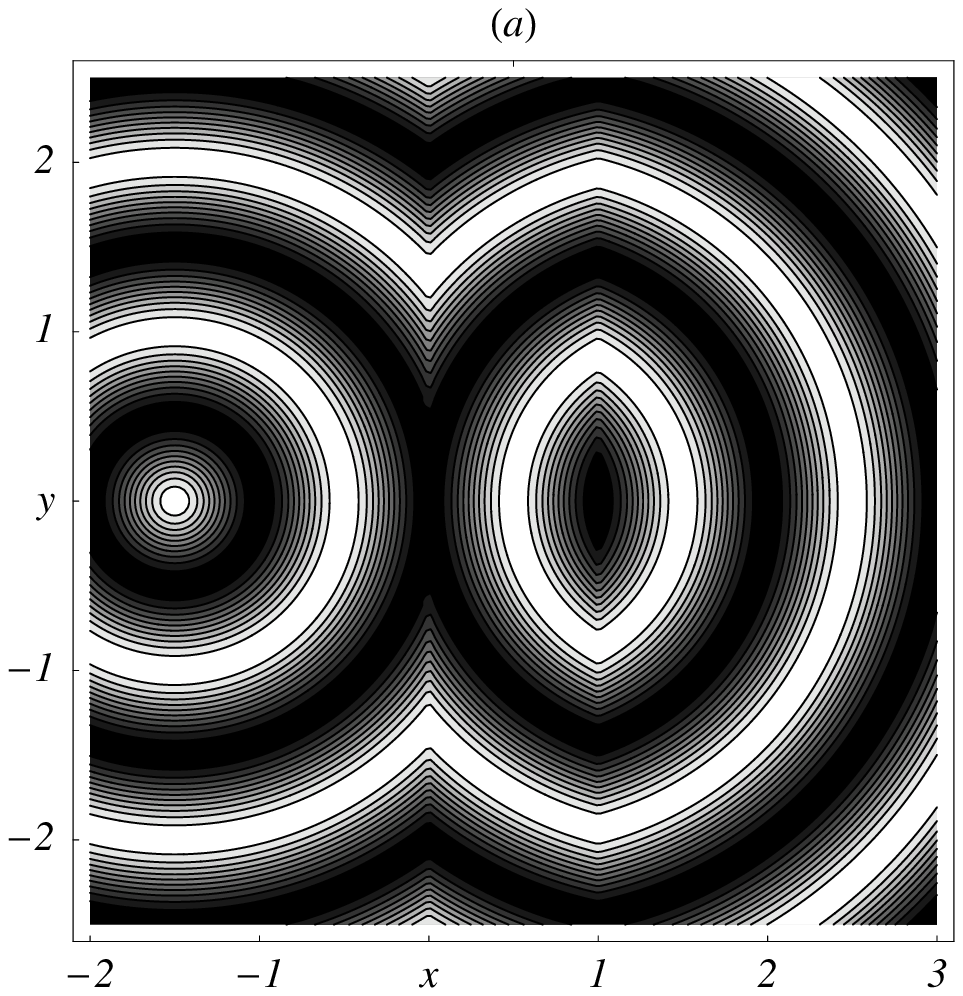}
\includegraphics[width=16.0pc]{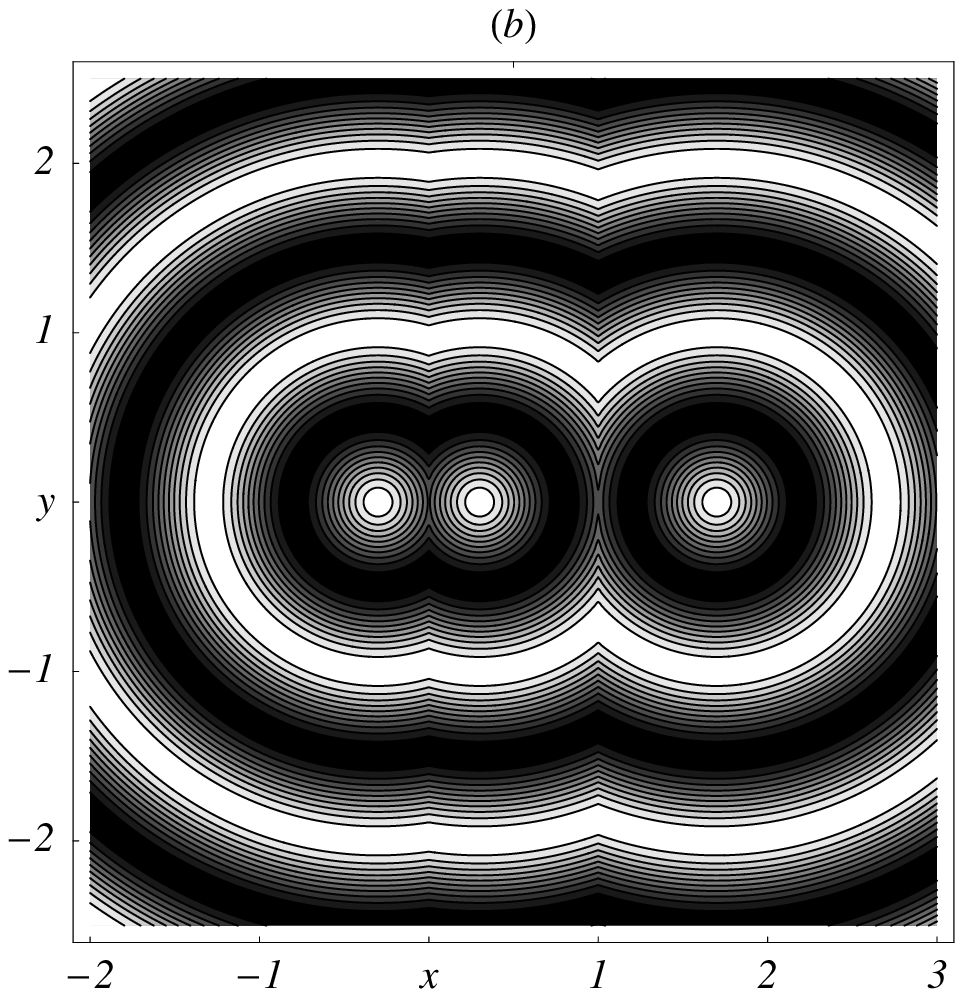}
\caption{
\small{
Propagation of electromagnetic waves in a perfect lens.
The lens facilitates the coordinate transformation
shown in Fig.\ \ref{fig:lens}.
Spherical waves in electromagnetic space 
are transformed into physical space. 
(a) The wave of Fig.\ \ref{fig:waves} (a) is emitted outside
the imaging range of the lens.
The wave is transformed by the lens, but the device is not
sufficiently thick to form an image.
(b) A wave is emitted inside the imaging range,
creating two images of the emission point, one inside the
device and one outside,
corresponding to the image points of Fig.\ \ref{fig:lens}. 
}}
\end{center}
\end{figure}

Suppose a transformation medium acts on the
electromagnetic field in the vacuum state.
Although the medium appears to merely transform
one nothing into another,
it may cause surprising physical phenomena:
for example \cite{Levitation}, 
the Casimir force \cite{Casimir}
may become repulsive in left-handed transformation media
\cite{LeoPhil,Veselago},
possibly even to the extent of levitating ultra-light metal foils.  
This effect \cite{Levitation} follows from the quantum optics 
of spatial transformation media set out in this paper. 

The quantum theory of light in isotropic media 
has been the subject of an extensive literature 
\cite{LeoReview,Media},
left-handed media have been considered \cite{KK}
with respect to their modification 
of the spontaneous emission of atoms,
but not transformation media in general. 
Such media are bi-anisotropic:
both the electric permittivity $\varepsilon$
and the magnetic permeability $\mu$ 
are real symmetric matrices that may spatially vary. 
Following
Kn\"oll, Vogel, and Welsch \cite{KVW} and
Glauber and Lewenstein \cite{GL},
we develop a concise quantum theory of light
in non-dispersive bi-anisotropic media.
Then we show how the quantum optics of transformation 
media emerges from this theory.
Although most transformation media are strongly dispersive, 
our simple non-dispersive theory is capable to describe the essence
of some unusual quantum effects,
for example quantum levitation due to repulsive Casimir forces
\cite{Levitation}.

\section{Bi-anisotropic media}

Consider quantum electromagnetism in non-dispersive media,
{\it i.e.} in frequency ranges where dispersion and absorption are
negligible. 
The media are described by 
the constitutive equations in SI units
\begin{equation}
\widehat{\mathbf{D}} = 
\varepsilon_0\varepsilon \widehat{\mathbf{E}} \,,\quad
\widehat{\mathbf{B}} =
\frac{\mu}{\varepsilon_0 c^2}\,\widehat{\mathbf{H}}
\,.
\label{eq:constitution}
\end{equation}
Throughout this paper we denote vectors in bold type.
The Hamiltonian, the quantum operator 
of the electromagnetic energy in such media, 
is \cite{LL8}
\begin{equation}
\hat{H} = \frac{1}{2}\int\left(
\widehat{\mathbf{E}} \cdot \widehat{\mathbf{D}} +
\widehat{\mathbf{B}} \cdot \widehat{\mathbf{H}}
\right) \mathrm{d}V
\,.
\label{eq:hamiltonian}
\end{equation}
Note that the Hamiltonian describes the energy of both 
the electromagnetic field and the medium.
In the absence of any external charges and currents 
\cite{KVW,GL},
we employ the operator $\widehat{\mathbf{A}}$
of the vector potential in order to describe the
electromagnetic field as
\begin{equation}
\widehat{\mathbf{E}} = 
- \frac{\partial  \widehat{\mathbf{A}}}{\partial t}
\,,\quad
 \widehat{\mathbf{B}} = \nabla \times  \widehat{\mathbf{A}}
\,.
\label{eq:fields}
\end{equation}
In Coulomb gauge \cite{GL,Weinberg},
\begin{equation}
\nabla \cdot \varepsilon \widehat{\mathbf{A}} = 0
\,,
\label{eq:coulomb}
\end{equation}
Maxwell's equations are naturally satisfied, except 
\begin{equation}
\nabla \times \widehat{\mathbf{H}} = 
\frac{\partial  \widehat{\mathbf{D}}}{\partial t}
\,.
\label{eq:maxwell}
\end{equation}
The quantum field $\widehat{\mathbf{A}}$
shall obey both the Coulomb and Maxwell
equations (\ref{eq:coulomb}) and (\ref{eq:maxwell}).
Additionally, we postulate the fundamental
equal-time commutation relation
\begin{equation}
\big[ \widehat{\mathbf{D}}(\mathbf{r}_1,t),
\widehat{\mathbf{A}}(\mathbf{r}_2,t) \big] =
i\hbar\,\delta^T(\mathbf{r}_1,\mathbf{r}_2) 
\label{eq:cr}
\end{equation}
where $\delta^T(\mathbf{r}_1,\mathbf{r}_2)$
denotes the transversal delta function
\cite{GL,Weinberg}
that is consistent with the Coulomb gauge  
(\ref{eq:coulomb}).
One may motivate the commutation relation (\ref{eq:cr})
by requiring that both the relation
$\widehat{\mathbf{E}} = 
- \partial\widehat{\mathbf{A}}/{\partial t}$
and the Maxwell equation (\ref{eq:maxwell})
should follow from the Hamiltonian (\ref{eq:hamiltonian}) 
in the form
\begin{equation}
\hat{H} =
\frac{1}{2\varepsilon_0}\int\left(
\widehat{\mathbf{D}}\, \varepsilon^{-1} 
\widehat{\mathbf{D}} +
\varepsilon_0^2 c^2
\widehat{\mathbf{B}}\, \mu^{-1} \widehat{\mathbf{B}}
\right) \mathrm{d}V 
\end{equation}
with $\widehat{\mathbf{B}} 
= \nabla \times  \widehat{\mathbf{A}}$.
Alternatively,
the canonical quantization procedure \cite{Weinberg}
of classical electromagnetism in media \cite{LL8}
leads to the commutation relation (\ref{eq:cr}).
Due to the interaction of the field 
with the dipoles of the medium,
the canonical field momentum depends on the medium;
and so relation (\ref{eq:cr}) connects the 
vector potential, a pure field quantity,
with the dielectric displacement
that describes both the field and the medium.

Since the Coulomb-Maxwell equations 
(\ref{eq:coulomb}) and (\ref{eq:maxwell})
are linear, we can expand 
$\widehat{\mathbf{A}}$
in terms of a complete set of modes
$\mathbf{A}_k(\mathbf{r},t)$
that obey the classical field equations
(\ref{eq:coulomb}) and (\ref{eq:maxwell}), 
\begin{equation}
\widehat{\mathbf{A}} = 
\sum_k \left(\mathbf{A}_k \hat{a}_k +
\mathbf{A}_k^* \hat{a}_k^\dagger
\right) \,.
\label{eq:modes}
\end{equation}
The mode functions  $\mathbf{A}_k(\mathbf{r},t)$
describe the classical aspects of the electromagnetic field,
for example electromagnetic wave packets
propagating in space and time through the medium,
whereas the operators $\hat{a}_k$ and $\hat{a}_k^\dagger$
are constant and describe the quantum excitations of 
the field, the photons.
In order to discriminate between photons with
different quantum numbers $k$,
we employ the scalar product
\begin{equation}
\left(\mathbf{A}_1, \mathbf{A}_2 \right) 
\equiv -\frac{i}{\hbar} \int \left(
\mathbf{A}_1^*\cdot\mathbf{D}_2 -
\mathbf{A}_2\cdot\mathbf{D}_1^*\right) \mathrm{d}V
\label{eq:scalar}
\end{equation}
where the $\mathbf{D}$ are connected to the 
$\mathbf{A}$ by the classical
constitutive equations (\ref{eq:constitution})
and (\ref{eq:fields}).
The modes may evolve in time, 
but their scalar product remains conserved,
because
\begin{eqnarray}
\frac{\mathrm{d} \left(\mathbf{A}_1, \mathbf{A}_2 \right) }
{\mathrm{d} t} 
&=& \frac{i}{\hbar} \int \Bigg(
\mathbf{E}_1^*\cdot\mathbf{D}_2 -
\mathbf{E}_2\cdot\mathbf{D}_1^*
- \mathbf{A}_1^*\cdot \frac{\partial\mathbf{D}_2}{\partial t} +
\mathbf{A}_2\cdot \frac{\partial\mathbf{D}_1^*}{\partial t} 
\Bigg)\, \mathrm{d}V \,,
\nonumber\\
\mathbf{A}\cdot \frac{\partial\mathbf{D}}{\partial t} 
&=& \mathbf{A}\cdot (\nabla\times\mathbf{H})
= - \nabla\cdot (\mathbf{A}\times\mathbf{H})
+\mathbf{B}\cdot\mathbf{H} \,,
\label{eq:adt}
\end{eqnarray}
together with the matrix symmetry of $\varepsilon$
and $\mu$ in the constitutive equations 
(\ref{eq:constitution}),
implies that the time derivative of 
$\left(\mathbf{A}_1, \mathbf{A}_2 \right)$ vanishes.
We require
\begin{equation}
\left(\mathbf{A}_k, \mathbf{A}_{k'} \right) = \delta_{kk'} 
\,,\quad
\left(\mathbf{A}_k^*, \mathbf{A}_{k'} \right) = 0
\,,
\label{eq:ortho}
\end{equation}
use these orthonormality conditions to extract the mode
operators from the expansion (\ref{eq:modes}) as
$\hat{a}_k =
\big(\mathbf{A}_k, \widehat{\mathbf{A}} \big)$
and
$\hat{a}_k^\dagger =
-\big(\mathbf{A}_k^*, \widehat{\mathbf{A}} \big)$,
and obtain from the fundamental field commutator 
(\ref{eq:cr}) the Bose commutation relations
\begin{equation}
[\hat{a}_k, \hat{a}_{k'}^\dagger] = \delta_{kk'} 
\,,\quad
[\hat{a}_k, \hat{a}_{k'}] = 0
\,.
\label{eq:bose}
\end{equation}
Like light in free space, 
light in a bi-anisotropic medium
consists of electromagnetic oscillators 
with the annihilation and creation operators 
$\hat{a}_k$ and $\hat{a}_k^\dagger$. 
The photons are the quanta of the independent 
degrees of freedom of the electromagnetic field.
As we have seen, their fundamental character is not
influenced by the medium.
However, 
the physics of the medium is important
in identifying the photons via the
scalar product (\ref{eq:scalar}),
and the mode functions $\mathbf{A}_k(\mathbf{r},t)$
naturally depend on the medium.

For stationary modes with frequencies $\omega_k$
the Hamiltonian (\ref{eq:hamiltonian})
appears as
\begin{equation}
\hat{H} = \frac{1}{2}\int\left(
\widehat{\mathbf{E}} \cdot \widehat{\mathbf{D}} +
\widehat{\mathbf{A}} \cdot 
\frac{\partial\widehat{\mathbf{D}}}{\partial t}
\right) \mathrm{d}V
=\sum_k \hbar\omega_k
\left(\hat{a}_k^\dagger\hat{a}_k + \frac{1}{2}\right)
\,,
\end{equation}
a sum of harmonic-oscillator energies,
including the zero-point energies $\hbar\omega_k/2$.
The frequencies $\omega_k$ depend on the 
medium and the boundary conditions.
The total zero-point energy $E_0$ is infinite,
but the derivative of $E_0$ with respect to
the boundaries is finite \cite{Casimir}:
giving rise to the Casimir force.

\section{Spatial transformation media}

Suppose that the electric permittivity $\varepsilon$
and the magnetic permeability $\mu$
in Cartesian coordinates $x^i$
are given by the matrices
\begin{equation}
\varepsilon = \frac{J J^T}{\mathrm{det} J}\, \varepsilon'
\,,\quad
\mu = \frac{J J^T}{\mathrm{det} J}\, \mu'
\,,\quad
J^i_j = \frac{\partial x^i}{\partial x'^j} \,.
\label{eq:epsmu}
\end{equation}
The coordinates $x^i$ and the scalars
$\varepsilon'$ and $\mu'$ are functions of 
some transformed coordinates $x'^j$.
In regions where the coordinates agree,
$J$ reduces to the unity matrix; 
and so $\varepsilon'$ and $\mu'$ directly describe
the dielectric properties of the medium here.
In the following
we show that the quantum optics of the 
general medium (\ref{eq:epsmu})
in physical space is mapped, 
in the primed coordinates,
to quantum electrodynamics
in the presence of a dielectric with $\varepsilon'$ and $\mu'$,
for the transformed vector potential
\begin{equation}
\widehat{\mathbf{A}}' = J \widehat{\mathbf{A}}  
\quad\mbox{or, equivalently,}\quad
\hat{A}_i' = \sum_j J^j_i \hat{A}_j
\,. 
\end{equation}
First we write $\varepsilon$ and $\mu^{-1}$ 
in component representation,
\begin{eqnarray}
\varepsilon &=& \pm\sqrt{\gamma}\,\gamma^{ij} \,\varepsilon' 
\,,\quad
\mu^{-1} = \pm \frac{\gamma_{ij}}{\sqrt{\gamma}\,\mu'}
\,,\quad
\gamma = \mathrm{det}(\gamma_{ij}) \,,
\nonumber\\
\gamma^{ij}
&=& \sum_{l}\frac{\partial x^i}{\partial x'^l}
\frac{\partial x^j}{\partial x'^l}
\,,\quad
\gamma_{ij} = \sum_{l}\frac{\partial x'^l}{\partial x^i}
\frac{\partial x'^l}{\partial x^j}
\label{eq:explicit}
\,.
\end{eqnarray}
These are the constitutive equations of spatial
transformation media \cite{LeoPhil}
expressed in terms of the spatial metric $\gamma_{ij}$,
the matrix inverse of $\gamma^{ij}$.
The $\pm$ sign distinguishes between right and left-handed 
coordinate systems where the Jacobian $\mathrm{det} J$
is positive or negative, respectively.
Left-handed systems thus correspond to left-handed materials
\cite{Veselago} where the eigenvalues of 
$\varepsilon$ and $\mu$ are negative \cite{LeoPhil}.
Consider for example the transformation  \cite{LeoPhil}
\begin{equation}
x' =
\left\{
\begin{array}{rcl}
x & \mbox{for} & x<0\\
-x & \mbox{for} & 0 \le x \le b\\
x-2b & \mbox{for} &  x>b
\end{array}
\right. 
\label{eq:left}
\end{equation}
illustrated in Fig.\ \ref{fig:lens}.
In the region $0 \le x \le b$,
Eq.\ (\ref{eq:left}) transforms a right-handed 
Cartesian coordinate system into a left-handed one.
We see from Eq.\ (\ref{eq:epsmu}) that
the transformation (\ref{eq:left})
is performed by a left-handed medium with
$\varepsilon=\mu=-1$ inside the slab 
and $\varepsilon=\mu=1$ outside
\cite{LeoPhil}.

Quantum electrodynamics in bi-anisotropic media
is characterized by the vector potential 
$ \widehat{\mathbf{A}}$ subject to both
the Coulomb gauge (\ref{eq:coulomb}) and
the Maxwell equation (\ref{eq:maxwell}).
We express the Coulomb-gauge condition as
\begin{equation}
0 = \pm \frac{\nabla\cdot\varepsilon
 \widehat{\mathbf{A}}}{\sqrt{\gamma}}
= \frac{1}{\sqrt{\gamma}}
\sum_{ij} 
\frac{\partial \sqrt{\gamma}\,\gamma^{ij}
\varepsilon'\hat{A}_j}{\partial x^i}
\,,
\label{eq:couco}
\end{equation}
which is the covariant form of the 3D divergence
\cite{Telephone}
if we interpret $\gamma_{ij}$ as the metric tensor
of physical space:
the physical Cartesian coordinates $x^i$ appear
as the back-transformed primed coordinates $x'^j$
of a space where the metric tensor $\gamma_{ij}'$
is the unity matrix;
the transformed space is flat.
Consequently, we obtain here
\begin{equation}
0 = \nabla'\cdot \varepsilon' \widehat{\mathbf{A}}'
\,.
\label{eq:couprime}
\end{equation}
Now we turn to the Maxwell equation (\ref{eq:maxwell}).
We apply the constitutive equation (\ref{eq:constitution})
for $\widehat{\mathbf{D}}$ with the electric permittivity 
(\ref{eq:explicit}) and write the electric field in terms
(\ref{eq:fields}) of the vector potential,
\begin{equation}
\varepsilon_0 \sum_j \gamma^{ij}
\varepsilon'\,
\frac{\partial^2\hat{A}_j}{\partial t^2}
= \mp
\frac{(\nabla\times\widehat{\mathbf{H}})^i}{\sqrt{\gamma}}
\,.
\label{eq:max1}
\end{equation}
Then we express the components of the curl 
in terms of the 3D Levi-Civita tensor $\epsilon^{ijl}$
for the spatial geometry characterized by the metric tensor
$\gamma_{ij}$ \cite{Telephone},
\begin{equation}
\pm\frac{(\nabla\times\widehat{\mathbf{H}})^i}{\sqrt{\gamma}}
= \sum_{jl}\epsilon^{ijl}\,
\frac{\partial\hat{H}_l}{\partial x^j}
\label{eq:curl}
\,.
\end{equation}
We invert the constitutive equations 
(\ref{eq:constitution}) for $\widehat{\mathbf{B}}$ 
using the representation (\ref{eq:explicit}) of $\mu^{-1}$
in terms of the metric tensor,
\begin{equation}
\hat{H}_i = \frac{\varepsilon_0 c^2}{\mu'}
\sum_j \gamma_{ij} 
\left(\pm\frac{\hat{B}^j}{\sqrt{\gamma}} \right) 
\,,
\end{equation}
apply the matrix $\gamma^{ij}$ to $\widehat{\mathbf{H}}$
and, like in Eq.\ (\ref{eq:curl}),
represent $\widehat{\mathbf{B}}$ as the curl of
$\widehat{\mathbf{A}}$ in covariant form
\cite{Telephone},
\begin{equation}
\sum_j \gamma^{ij} \hat{H}_j = 
\frac{\varepsilon_0 c^2}{\mu'}
\sum_{jl}\epsilon^{ijl}\,
\frac{\partial\hat{A}_l}{\partial x^j}
\,.
\label{eq:max2}
\end{equation}
Equations (\ref{eq:max1}), (\ref{eq:curl}), and (\ref{eq:max2})
combined are covariant, 
such that we obtain in transformed space
\begin{equation}
\frac{1}{\varepsilon'}
\nabla'\times
\frac{1}{\mu'}\nabla'
\times\widehat{\mathbf{A}}'
= -\frac{1}{c^2}\,
\frac{\partial^2\widehat{\mathbf{A}}'}{\partial t^2}
\label{eq:maxprime}
\,.
\end{equation}
Consequently,
the Coulomb-Maxwell equations 
(\ref{eq:couprime}) and (\ref{eq:maxprime})
for the transformed 
quantum field $\widehat{\mathbf{A}}'$ are the ones
of flat space
filled with a medium of permittivity $\varepsilon'$
and permeability $\mu'$.
All the other building blocks of the
quantum theory of light in transformation media,
the Hamiltonian (\ref{eq:hamiltonian}),
the field commutator (\ref{eq:cr}),
and the scalar product (\ref{eq:scalar}),
are naturally covariant
and are therefore mapped to flat space, too.

\section{Conclusions}

We developed the quantum theory of light in dispersion-less
bi-anisotropic media.
The theory shows how transformation media \cite{PSS,LeoPhil}
map quantum optics in physical space
to the quantum electromagnetism in flat space.
Note that
even in the extreme case when the 
electromagnetic field is in the vacuum state,
when literally nothing appears to be mapped to nothing,
transformation media may cause surprising physical effects:
the Casimir force in left-handed materials
\cite{Veselago} may be repulsive
\cite{Levitation}.

We thank J. W. Allen, M. Killi, J. B. Pendry and S. Scheel 
for their comments
and the Leverhulme Trust for financial support.

\end{document}